# Discovery of Romanov V20, an Algol-Type Eclipsing Binary in the Constellation Centaurus, by Means of Data Mining

## Filipp Dmitrievich Romanov

*ORCID: 0000-0002-5268-7735; Moscow, Russian Federation; filipp.romanov.27.04.1997@gmail.com*

**Abstract** I report my discovery of the large-amplitude Algol-type eclipsing binary system which was initially added to the AAVSO International Variable Star Index (VSX) under the designation of Romanov V20. I describe selection criteria for searching for variability among other stars, the search of photometric data from several sky surveys, and my observations using remote telescopes, and the analysis of the data in the VStar software. I find the orbital period, eclipse duration, and magnitude range in Johnson B, V and Sloan g, r, i bands for primary and secondary eclipses.

## 1. Introduction

During the course of analysis of the AllWISE catalog (Cutri *et al.* 2014), I discovered a new variable star. I checked that the star was not previously known as a variable in the AAVSO VSX, in the VizieR catalogues, or in the SIMBAD Astronomical Database. Using the publicly available All-Sky Automated Survey for Supernovae (ASAS-SN) Sky Patrol (Shappee *et al.* 2014; Kochanek *et al.* 2017) data, I found the period of eclipses and magnitude range in Johnson V band (including magnitude of secondary eclipse) and duration of primary eclipse. The star was added to the AAVSO VSX (Watson *et al.* 2006) on 03 December 2018 under the designation of Romanov V20. Later its variability was confirmed in the *ASAS-SN catalogue of variable stars – V. Variables in the Southern hemisphere* (Jayasinghe *et al.* 2020) under the name of ASASSN-V J112124.71-522143.6.

In this paper, I describe my method of search which led to the discovery of this star and my refinement of the initially determined parameters of the system, in part on the basis of multicolor photometry from remote telescopes.

## 2. Information about Romanov V20

Its position (epoch J2000.0) according to Gaia DR2 (Gaia Collaboration *et al.* 2018) is: R.A. 11h 21m 24.68s, Dec. –52° 21' 43.7"; galactic coordinates: 289.2744°, +8.1061° (Centaurus).

Other names of this star include:
2MASS J11212468-5221437 =
WISEA J112124.66-522143.7 =
GALEX J112124.6-522143 =
GSC 08225-00671 =
UCAC4 189-058773 =
USNO-B1.0 0376-0339496.

Table 1 presents the data about the star: magnitudes and colors from several catalogs and the mean (calculated by the author) of maximum B and V magnitudes from the APASS DR10 (Henden *et al.* 2018) Epoch Photometry Database. The rounded value of the geometric and photogeometric distance posteriors (Bailer-Jones *et al.* 2021) for this star is 1550 pc. The effective temperature (from Gaia Data Release 2) is 8020 K.

## 3. Selection criteria for searching for variability

This star was found in the AllWISE catalog using the TAP VizieR service (http://tapvizier.u-strasbg.fr/adql/). In this catalog, objects have a variability flag value (from 0 to 9) which indicates the probability that the source flux measured on the individual WISE exposures is variable; values > 7 have the highest probability of being true variables. But, because the catalog does not contain information about the detected variable stars, such as classifications or periods, these objects are not known as variables on this basis alone. The text of my request using Astronomical Data Query Language (Osuna *et al.* 2008) is attached in Appendix A. It was designed to exclude (by color indices, with a large margin) red variable stars, such as semiregular or Mira-type, while the magnitude was limited to W1 < 13 to exclude stars that would be too faint to observe with an amateur telescope. The search area was chosen within several degrees of the Galactic plane. From the obtained table, I selected those stars that have the variability flag 9 for the W1 and W2 bands, and checked them for variability in the data of the ASAS-SN Sky Patrol.

## 4. Observations

After adding this star to the AAVSO VSX, I used the ephemeris given in VSX to calculate times of primary minimum, and I observed the primary eclipse using the remote telescope T32 (0.43-m f/6.8 reflector + Charge-Coupled Device) of iTelescope.Net at Siding Spring Observatory, Australia. Seventeen photos with 60 seconds exposure time and Johnson V filter were obtained on 26 February 2020, but only part of the duration of the primary eclipse was recorded due to weather conditions. For photometric measurements I used MAXIM DL Pro Version 6.23 Demo software (Diffraction Limited 2020) and the AAVSO star chart. These values are presented in Table 2. Figure 1 shows a finding chart for this variable star, created from images taken at this remote telescope during this observing run.

Table 1. Magnitudes and colors of Romanov V20.

| Source | Magnitude | Color Index |
|---|---|---|
| 2MASS (Two Micron All-Sky Survey) | J = 12.86; H = 12.65; K = 12.60 | J–K = 0.26 |
| AllWISE (Wide-field Infrared Survey Explorer) | W1 = 12.63; W2 = 12.69 | W1–W2 = -0.06 |
| APASS DR10 (AAVSO Photometric All-Sky Survey) | V = 13.46; B = 13.66 | B–V = 0.20 |
| Gaia DR2 | G = 13.43; BP = 13.54; RP = 13.19 | BP–RP = 0.35 |
| GALEX GR6 (Galaxy Evolution Explorer) | FUV = 19.57; NUV = 17.09 | FUV–NUV = 2.48 |

The magnitudes (from APASS DR10) and positions (from Gaia DR2) of the comparison stars marked in Figure 1 are shown in Table 3.

On 18 January 2021, I observed another primary eclipse on T32. I obtained 10 V-band images, but I still did not detect the moment of minimum brightness. Therefore, I requested observations of eclipses of this variable star on the remote telescopes of AAVSOnet (Simonsen 2011). From 12 March to 27 March 2021, images of Romanov V20 were taken with the Johnson V, B, and Sloan r, i filters on AAVSOnet telescopes OC61 (Optical Craftsman 0.61-m telescope located at Mount John University Observatory, New Zealand) and BSM Berry2 (72-mm refractor of Bright Star Monitor Station located in Perth, Australia). Exposures were 60 seconds for imaging with all filters except B (120 seconds).

I used AAVSO VPHOT (online tool for photometric analysis; AAVSO 2021) for my photometric measurements from these images. The magnitudes of the comparison stars were the same as shown in Table 3 (I used r' and i' magnitudes of stars for comparison for images taken with r and i filters, because I ignored minor differences between these magnitude systems). As a result, I obtained 1964 values of brightness of Romanov V20. The purpose of obtaining images from the AAVSOnet remote telescopes was not to clarify the time of eclipses, but to measure the depths of the eclipses in different filters.

I uploaded all my photometric measurements to the AAVSO International Database (AID).

## 5. Data analysis

In the beginning, I analyzed the first data from iTelescope T32 and the sky surveys data. I made a heliocentric correction for the times of my photometric values and downloaded the photometric data of this star from the following sky surveys: ASAS-SN Johnson V (from 04 February 2016 to 03 August 2018) and Sloan g (from 13 June 2018 to 10 August 2020) bands, APASS V and g' bands (from 19 February 2011 to 01 June 2014), and All Sky Automated Survey: ASAS-3 (Pojmański 2002) V band (from 07 December 2000 to 30 July 2009). ASAS-3 observations are assigned four quality flags, from A to D (in order of decreasing quality). I used epoch photometric data (HJD-2450000) from the MAG_0 column with the quality grades A and B for the analysis.

Table 2. The first results of photometric measurements of magnitude of Romanov V20.

| Time (JD) | Magnitude (V) | Error |
|---|---|---|
| 2458906.249468 | 14.571 | 0.010 |
| 2458906.250637 | 14.571 | 0.010 |
| 2458906.251667 | 14.510 | 0.010 |
| 2458906.252778 | 14.531 | 0.009 |
| 2458906.253854 | 14.455 | 0.009 |
| 2458906.254919 | 14.443 | 0.009 |
| 2458906.256019 | 14.427 | 0.009 |
| 2458906.257072 | 14.385 | 0.009 |
| 2458906.258137 | 14.339 | 0.009 |
| 2458906.259225 | 14.318 | 0.009 |
| 2458906.260266 | 14.303 | 0.008 |
| 2458906.261331 | 14.265 | 0.008 |
| 2458906.262407 | 14.182 | 0.008 |
| 2458906.263495 | 14.220 | 0.008 |
| 2458906.264618 | 14.173 | 0.008 |
| 2458906.265683 | 14.135 | 0.008 |
| 2458906.266759 | 14.148 | 0.008 |

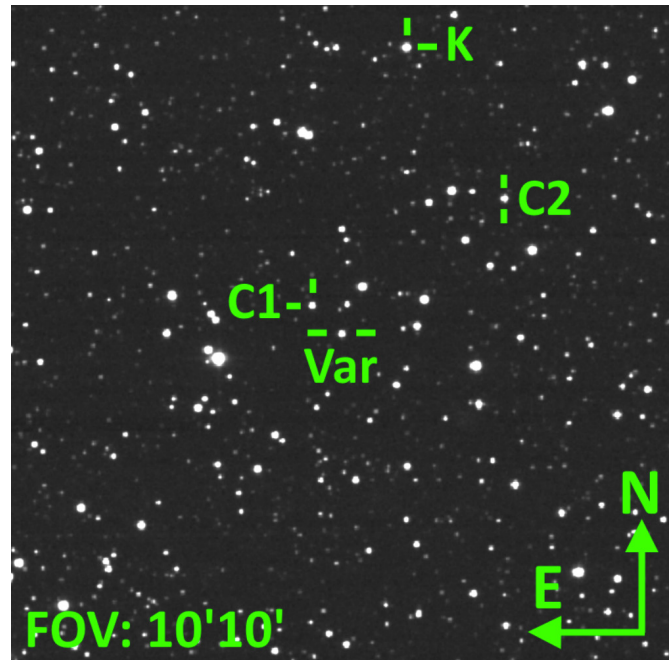

Figure 1. Finding chart for Romanov V20. Var is Romanov V20, C1 and C2 are comparison stars, K is check star.

I analyzed all data in the VSTAR software (Benn 2012). As a result, I improved the information previously added by me to VSX: period, magnitude range, and duration of eclipses. The fact that the secondary eclipse is at phase 0.5 suggests that the orbit is circular. These results are shown in Table 4. The resulting combined phased light curve of primary eclipse is

Table 3. Comparison stars.

| Ident. | Name | R.A. (J2000.0) h m s | Dec. (J2000.0) ° ' " | B | V | g' | r' | i' | B–V |
|---|---|---|---|---|---|---|---|---|---|
| C1 | UCAC4 189-058786 | 11 21 27.64 | –52 21 18.07 | 15.32 ± 0.046 | 14.57 ± 0.14 | 14.87 ± 0.023 | 14.33 ± 0.131 | 14.18 ± 0.225 | 0.75 |
| C2 | UCAC4 189-058714 | 11 21 08.59 | –52 19 38.75 | 14.60 ± 0.039 | 13.96 ± 0.131 | 14.22 ± 0.033 | 13.74 ± 0.119 | 13.59 ± 0.208 | 0.64 |
| K | UCAC4 189-058753 | 11 21 18.49 | –52 17 21.02 | 13.50 ± 0.03 | 12.98 ± 0.118 | 13.19 ± 0.029 | 12.77 ± 0.114 | 12.69 ± 0.212 | 0.52 |

Table 4. Parameters of Romanov V20.

Period: 1.267496 days
Epoch of primary eclipse 2457934.052 HJD
Duration of primary eclipse: 15% (4.56 hours).

| Band | Max | Min I | Min II | Source |
|------|-----|-------|--------|--------|
| B | 13.65 ± 0.03 | 16.7 ± 0.2 | 13.73 ± 0.03 | AAVSOnet |
| V | 13.45 ± 0.02 | 16.45 ± 0.2 | 13.54 ± 0.02 | ASAS-SN; AAVSOnet |
| g | 13.52 ± 0.02 | 16.6 ± 0.1 | 13.59 ± 0.02 | ASAS-SN |
| r | 13.37 ± 0.03 | 15.95 ± 0.15 | 13.49 ± 0.02 | AAVSOnet |
| i | 13.45 ± 0.04 | 15.65 ± 0.15 | 13.61 ± 0.02 | AAVSOnet |
| W1 | 12.50 ± 0.02 | 13.6 ± 0.1 | 12.97 ± 0.05 | NEOWISE-R |
| W2 | 12.55 ± 0.02 | 13.85 ± 0.15 | 13.05 ± 0.1 | NEOWISE-R |

shown in Figure 2. Figure 3 shows the combined phase plot for secondary eclipse in the ASAS-SN V and g data. Figure 4 shows phased light curve from the AAVSOnet data (V, B, r, i bands).

After finding these parameters of the system, I also created the phase plot from the NEOWISE-R data (Mainzer *et al.* 2011): I downloaded data (from 12 January 2014 to 21 June 2019) in the W1 and W2 bands from the NASA/IPAC Infrared Science Archive (NASA/IPAC 2020) from NEOWISE-R Single Exposure (L1b) Source Table. Figure 5 shows the HJD phased light curve plotted with the VSTAR software from these data. Table 4 shows the range of variability in all the observed bands.

The fact that the eclipse depths change with passband, with the primary eclipse being deeper in the B (AAVSOnet data) and g bands (ASAS-SN data) and the secondary eclipse being much deeper in the longer-wavelength bands (AAVSOnet Sloan r and i, and NEOWISE-R W1 and W2), may be explained if the hotter component of the system (judging by the color indices, this is a star of spectral class A or F) is eclipsed by the cooler one (which has relatively brighter magnitudes in the infrared range than in optical) and vice versa.

## 6. Conclusions

I used both my own photometric measurements in various filters and data from sky surveys to determine the basic parameters of a newly discovered eclipsing binary, Romanov V20 = ASASSN-V J112124.71-522143.6. I conclude that the data from sky surveys are quite enough to determine the period, duration, and epoch of eclipses for bright eclipsing stars.

I showed that an amateur astronomer, who does not have astronomical equipment, but only has a personal computer and access to the Internet, can both search and discover variable stars based on open photometric data from sky surveys, and can research variable stars using observations with remote telescopes. This is a valuable contribution to the science of astronomy; moreover, in the future, such variable stars may become objects of professional astronomical research.

## 7. Acknowledgements

I am grateful to the AAVSO for granting me a complimentary membership for 2021 (thus giving me access to AAVSOnet,

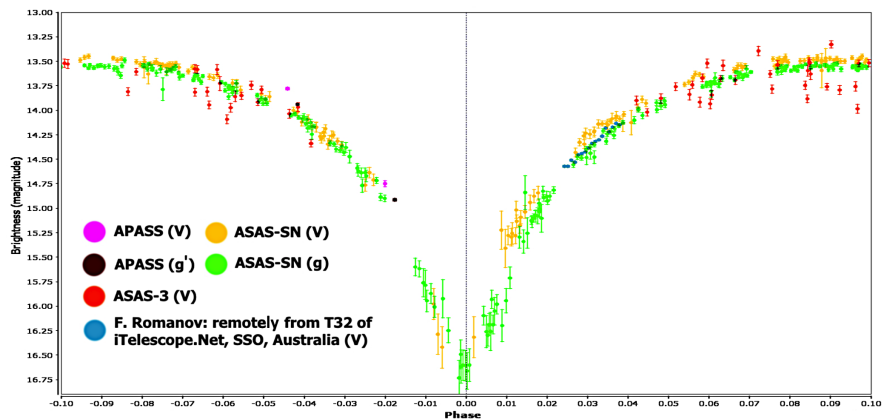

Figure 2. Phase plot for Romanov V20 (primary eclipse).

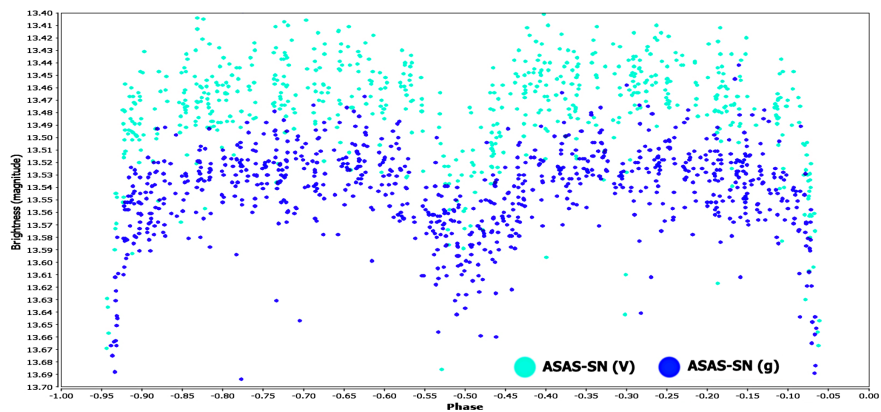

Figure 3. Phase plot for Romanov V20 (secondary eclipse).

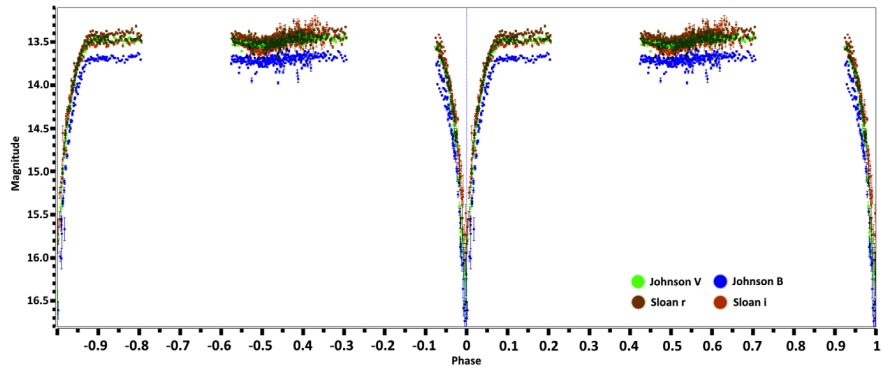

Figure 4. Combined phase plot from the AAVSOnet data.

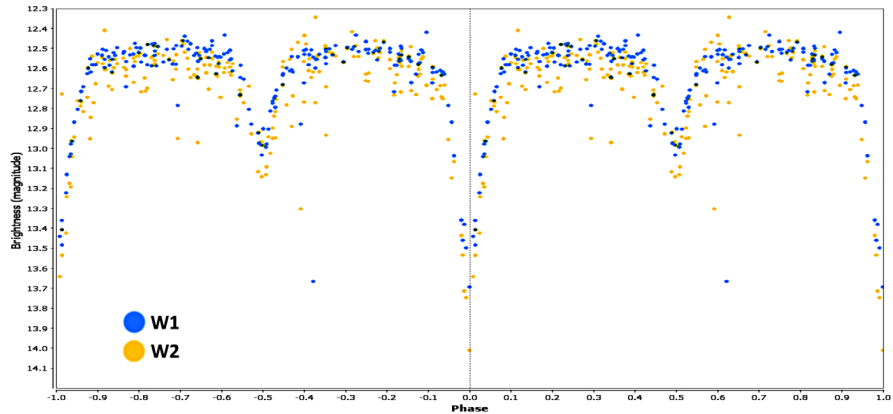

Figure 5. Phase plot for Romanov V20 (based on the NEOWISE-R data).

VPhot, and APASS Epoch Photometry data) and for imaging Romanov V20 with AAVSOnet remote telescopes after the approval of my proposal #184.

Also, I am grateful to iTelescope.Net for giving me some complimentary points for observing (including for imaging Romanov V20) with their remote telescopes.

This publication makes use of data products from the Two Micron All Sky Survey, which is a joint project of the University of Massachusetts and the Infrared Processing and Analysis Center/California Institute of Technology, funded by the National Aeronautics and Space Administration and the National Science Foundation.

This research has made use of the VizieR catalogue access tool, CDS, Strasbourg, France (DOI: 10.26093/cds/vizier). The original description of the VizieR service was published in *Astron. Astrophys.*, **143**, 23.

This paper is based in part on observations made with the Galaxy Evolution Explorer (GALEX). GALEX is a NASA Small Explorer, whose mission was developed in cooperation with the Centre National d'Etudes Spatiales (CNES) of France and the Korean Ministry of Science and Technology. GALEX is operated for NASA by the California Institute of Technology under NASA contract NAS5-98034.

This research has made use of the NASA/IPAC Infrared Science Archive, which is funded by the National Aeronautics and Space Administration and operated by the California Institute of Technology.

**Appendix A: Query using Astronomical Data Query Language to extract WISE data for this research.**

```
-- output format : text
SELECT "II/328/allwise".AllWISE, "II/328/allwise".RAJ2000,
    "II/328/allwise".DEJ2000,  "II/328/allwise".W1mag,
    "II/328/allwise".W2mag,  "II/328/allwise".var
FROM "II/328/allwise"
WHERE 1=CONTAINS(POINT('ICRS',"II/328/allwise".
    RAJ2000,"II/328/allwise".DEJ2000), BOX('ICRS',
    170.00000, -52.00000, 5., 5.))
AND W1mag-W2mag<0 AND Jmag-Kmag<0.3 AND W1mag
    <13
```